\documentclass[prl,twocolumn,superscriptaddress,showpacs,epsf]{revtex4}
\usepackage{rotating}
\usepackage{graphicx}
\usepackage{latexsym}
\usepackage{amssymb}
\usepackage{amsfonts}
\usepackage{soul}
\usepackage{color}
\begin{document}
\bibliographystyle{apsrev}
\title{On the origin of the reversed vortex ratchet motion}

\author{W. Gillijns, A.V. Silhanek, and V. V. Moshchalkov}
\affiliation{INPAC - Institute for Nanoscale Physics and
Chemistry, K.U.Leuven, Celestijnenlaan 200D, B--3001 Leuven,
Belgium}
\author{C.J. Olson Reichhardt and C. Reichhardt}
\affiliation{Theoretical Division, Los Alamos National Laboratory,
Los Alamos, New Mexico 87544, USA}


\date{\today}
\begin{abstract}
We experimentally demonstrate that the origin of multiply reversed rectified vortex motion in an asymmetric pinning landscape is a consequence not only of the vortex-vortex interactions but also essentially depends on the ratio between the characteristic interaction distance and the period of the asymmetric pinning potential.
Our system consists of an Al film deposited on top of a square array of {\it size-graded} magnetic dots with a constant lattice period $a=2~\mu$m. Four samples with different periods of the size gradient $d$ were investigated. For large $d$ the dc voltage $V_{dc}$ recorded under a sinusoidal ac excitation indicates that the average vortex drift is from bigger to smaller dots for all explored positive fields. As $d$ is reduced a series of sign reversals in the dc response are observed as a function of field. We show that the number of sign reversals increases as $d$ decreases. These findings are in agreement with recent computer simulations and illustrate the relevance of the different characteristic lengths for the vortex rectification effects.
\end{abstract}


\maketitle


When independent overdamped particles in an asymmetric potential are subject to an oscillatory driving
force with zero mean, a stationary flux of particles can be achieved. This rectified motion of particles, known as a rocked ratchet, is basically the result of the broken spatial symmetry of the potential energy\cite{reviews}. It has been recently shown that this effect can be used to manipulate the motion of flux quanta in superconductors with asymmetric pinning landscapes\cite{lee,villegas,clecionature,clecioprl,silhanekapl}. In this case, the fact that the particles (flux lines) cannot be regarded as independent entities leads to a far richer ratchet motion which can even reverse the direction of the average motion of vortices.

The relevance of the vortex-vortex interaction has already been  pointed out by Souza Silva {\it et al.}\cite{clecionature} who describe the sign reversal of the ratchet signal as a result of trapping several vortices in each pinning site. Lu {\it et al.}\cite{lu} predicted that reversed ratchet motion should also be present in a simpler system consisting of one dimensional asymmetric pinning potential. In the former description the penetration depth $\lambda$ was larger than the period of the ratchet potential $d$, whereas in the latter model $\lambda = d$. Similarly, this effect has been observed  experimentally \cite{shalom} and modelled theoretically \cite{marconi} for arrays of Josephson junctions with asymmetric periodic pinning, where $\lambda = \infty$. These studies have unambiguously identified
a strong vortex-vortex interaction as
the basic ingredient for producing ratchet reversals.
This situation is achieved when the intervortex distance $a_0$ becomes smaller than $\lambda$. Although the inequality $a_0 < \lambda$ seems to represent a {\it necessary} condition, it cannot be sufficient since the natural scale where the inversion symmetry is broken, $d$, is not explicitly taken into account. An important question is what the ultimate conditions are for obtaining a reversed ratchet motion and how the number of sign reversals depends on the characteristic scales of the system.

In this work we directly address these questions by investigating the vortex ratchet motion in samples with different periods of the asymmetric pinning potential $d$. For $d>\lambda$ no sign reversal is detected. This finding demonstrates that $\lambda > d$ represents a prerequisite to obtain reversed ratchet motion. In addition we show that as $d$ is decreased the number of sign reversals increases. We argue that this effect can be qualitatively described within a collective pinning scenario.

\vspace{0.5cm}
\begin{figure}[b]
\begin{center}
\end{center}
\caption{ (color online) Scanning electron microscopy images of the square array of magnetic dots with a size-graded period of (a) 162 $\mu$m, (b) 34 $\mu$m, (c) 18 $\mu$m, and (d) 10 $\mu$m. In panel (a) the positive current orientation to obtain a positive ratchet signal with positive field is indicated (see panel (a) in Fig.~\ref{fig2}). The vertical bars indicate a 40 $\mu$m scale.} \label{fig1}
\end{figure}

The pinning potential is created by a square array (period $a=2 \mu$m) of ferromagnetic dots of a size that changes linearly from
$1.2 \mu$m to $0.4 \mu$m over a distance $d$ (see Fig.~\ref{fig1}).
Four different periods $d = 10$, 18, $34$, and 162 $\mu$m were used. The magnetic dots are [0.4 nm Co/1.0 nm Pt]$_{n}$ multilayers with $n=10$ grown on top of a 2.5 nm Pt buffer layer. The dot arrays are then covered with a 5 nm thick Si layer followed by an Al bridge of thickness $t=50$ nm. In this way the Al is insulated from the ferromagnetic template and proximity effects are suppressed.

Since the typical field excursion used to explore the superconducting state is much smaller than the coercive field of the Co/Pt multilayers, the magnetic dots remain in the as-grown state during our measurements. The virgin state of the dots actually consists of a multidomain structure with zero net magnetization \cite{gillijns-PRB,silhanek-milosevic}. Still the stray field generated by these domains is enough to deplete the superconducting condensate locally on top of the dots, making the dots behave like pinning centers


\begin{figure*}[t]
\centering
\begin{minipage}[c]{0.98\linewidth}
\centering
\end{minipage}
\hfill
\centering \caption{(color online) Rectified signal $V_{dc}$ as a function of field $H$ and temperature $T$ for samples with period of the asymmetric pinning potential (a) 162 $\mu$m, (b) 34 $\mu$m, (c) 18 $\mu$m, and (d) 10 $\mu$m corresponding to the images shown in Fig.~\ref{fig1}. In all cases the external ac current is applied along the $x$-axis. The horizontal dashed lines indicate the temperature at which $\lambda \sim d$.} \label{fig2}
\end{figure*}

The electrical transport measurements were carried out in a He4 cryostat with a base temperature of 1 K and a stability better than 100 nK, which is mounted with standard electronics. The normal/superconductor (N/S) phase transition (not shown) for all studied samples shows a featureless and nearly linear $T_{c}(H)$ phase boundary. The slope of these N/S boundaries corresponds to a superconducting coherence length at zero temperature $\xi(0) \sim 114 \pm 5$ nm. From the dirty limit expression $\xi(0) \sim 0.855 \sqrt{\xi_0\ell}$, where $\xi_0 \sim 1400$ nm is the BCS coherence length for Al and $\ell$ the electronic mean free path, we estimate $\ell \sim 13$ nm. Taking into account the finite thickness of the film and the reduced mean free path we obtain an effective penetration depth $\lambda(0) \sim 1.46~\mu$m. On the right axes of Fig.~\ref{fig2} we have indicated the variation of $\lambda$ with temperature. Within the temperature window of our measurements $\lambda$ is always bigger than the vortex-vortex separation $a_0=\sqrt{\phi_0/H}$, where $\phi_0$ is the flux quantum\cite{comment}. This indicates that the {\it vortex-vortex interaction is in all cases significant in our experimental conditions}.

Rectification effects were investigated by applying a sinusoidal excitation of $500~\mu$A amplitude and frequency $f=1$ kHz while simultaneously recording the average dc voltage with a nanovoltmeter. The acquired dc signal, when the current is parallel to the gradient, i.e. Lorentz force $F_L$ parallel to the iso-size lines ($x$-axis in Fig.~\ref{fig1}), was below our experimental resolution (not shown). This is consistent with the fact that for this direction of the Lorentz force the system is fully symmetric. A different situation is expected when the external ac current shakes the vortex lattice with a $F_L$ in the same direction as the gradient ($y$-axis in Fig.~\ref{fig1}).  In this case a flux line feels an effective pinning potential $U_p$
schematically shown in Fig.~\ref{fig3}, with asymmetric forces $f=-\partial U_p/\partial y$ for motion along $+y$ and $-y$ orientations. The lack of the inversion symmetry of $U_p(y)$ is a natural consequence of the increasing pinning strength as the radius of the dots becomes larger \cite{gillijns-PRB,gillijns-unp}. Under these circumstances an external ac excitation should induce a net motion of vortices in the direction of the smaller slope of $U_p$, i.e. towards $+y$. This is indeed confirmed by the positive dc voltage $V_{dc}$ at positive fields obtained for the largest period of size-gradient ($d/a=81$), shown in Fig.~\ref{fig2}(a). For this sample, the ratchet signal is strong only when the density of dots outnumbers the density of vortices, i.e. for fields below the first matching field of the square array $H_1^{2D}=\phi_0/a^2=0.51$ mT.
Near $H_1^{2D}$, the hopping of vortices is suppressed
since most of the pinning centers are occupied,
and therefore the ratchet signal decreases. The most important feature of this figure is the lack of reversal ratchet in the whole range of fields studied. We argue that this behavior, present even though the vortex-vortex interaction is important, is due to the large period of the asymmetric pinning landscape $d$ in comparison with the penetration depth $\lambda$. For the sake of clarity, in Fig.~\ref{fig2} we added horizontal dashed lines at the temperatures where $\lambda \sim d$.

\begin{figure}[h]
\begin{center}
\end{center}
\caption{(color online) Schematic representation of the asymmetric ratchet potential $U_p$ along the $y$-axis (see also Fig.~\ref{fig1}). The smaller average slope when vortices move towards $+y$ is the easy direction for the net vortex drift under oscillatory excitations.} \label{fig3}
\end{figure}

The most conclusive evidence that the ratio $d/\lambda$ is indeed the relevant parameter which properly accounts for the appearance of reversed ratchet motion comes from Fig.~\ref{fig2}(b),(c), and (d) where the ratchet period is progressively reduced to meet the condition $d<\lambda$. In contrast to the previous sample, where $d/\lambda>1$, for all these samples clear multiple sign reversals of the ratchet signal are observed. {\it This finding shows that the existence of a strong vortex-vortex interaction is not a sufficient condition to reverse the easy ratchet direction, but instead is the relative size of $\lambda$ in comparison with the period of the ratchet potential $d$ which determines the occurrence of this effect}. Furthermore, the number of sign reversals in the same window of temperature and field ranges from none for $d/a=81$, 2 for $d/a=17$, 3 for $d/a=9$, and to 5 for $d/a=5$, illustrating the influence of this ratio on the number of observed reversals of sign of the vortex rectification effect.

\begin{figure}[h]
\begin{center}
\end{center}
\caption{Simulated ratchet response $\langle V\rangle$ vs
vortex density $n_v$
for systems with different pinning periods (a) $d=8\lambda$, (b) $d=3\lambda$,
(c) $d=\lambda$, and (d) $d=0.5\lambda$.}
\label{fig4}
\end{figure}

Lu {\it et al.}\cite{lu} recently showed that a two
dimensional vortex lattice interacting with a one-dimensional
ratchet potential produces a conventional ratchet when the
vortex lattice is highly ordered in such a way that the vortex-vortex
interactions cancel. In contrast, disordered configurations of the
vortex lattice lead to reversed ratchet motion.
The vortex lattice disordering occurred due to buckling
transitions of the vortices confined in each row of the ratchet
over the field range considered in Ref.~\cite{lu},
but additional reversals are expected at lower fields.
As shown in Ref.~\cite{Levitov}, a two-dimensional vortex
lattice interacting with a one-dimensional periodic pinning array undergoes
a series of continuous phase transitions as the magnetic field is decreased
provided that the vortex-vortex interaction remains strong enough to produce a
triangular ordering of the vortex lattice.
The number of possible vortex lattice orderings that appear for decreasing
vortex density increases when the lattice constant of the pinning array
decreases, so we expect to find more low-field
order-disorder transitions, and thus
more low-field ratchet reversals, for smaller lattice periods $d$.
To illustrate this, we have performed new simulations with the system
described in Ref.~\cite{lu} for pinning strength
$A_p=0.15f_0$, ac field $F_{ac}=0.065f_0$
and ratchet periods $d=8\lambda$, $3\lambda$, $\lambda$, and $0.5\lambda$,
where $f_0=6\phi_0^2/2\pi\mu_0\lambda^3$.
In Fig.~\ref{fig4} we plot the net velocity $\langle V\rangle$ versus
vortex density $n_v$,
showing that
the number of ratchet reversals for $n_v<1.1$  changes
from no reversals at $d=8\lambda$ to two reversals for $d=3\lambda$,
four at $d=\lambda$, and more than eight reversals at $d=0.5\lambda$.
This demonstrates that as the ratchet period $d$ decreases, more ratchet
reversals occur, in agreement with the experimental results.
We note that our experimental system represents, for fields $H<H_1^{2D}$,
a discrete two dimensional version of the situation described by
the simulations.

As a last remark we would like to point out that strictly speaking it is unlikely that the single parameter $d/\lambda$ properly describes the crossover from no-reversal to reversed ratchet. For instance a correlation length similar to the Larkin-Ovchinnikov\cite{LO,garten} (LO) correlation length $R_c=\sqrt{c_{66}a_0/F_p}$, where $c_{66}$ is the shear elastic modulus of the vortex lattice and $F_p$ is the pinning force density, should be taken into account instead of $\lambda$. Unfortunately the previous scale is a theoretical estimate for thick samples with weak random pinning and to our knowledge an extension of this equation to include thin film geometries with periodic array of pinning centers is still not available in the literature.

Concluding, we have experimentally demonstrated that when the characteristic vortex-vortex interaction scale length $\lambda$ is of the order of the period of the asymmetric pinning landscape $d$ an inversion of the effective ratchet potential for vortex motion occurs. This implies that the inequality $d < \lambda$ which represents a sufficient condition for the occurrence of multiple reversal in vortex systems should remain valid for any ratchet system with interacting particles such as colloids and granular materials.

This work was supported by the K.U.Leuven Research Fund
GOA/2004/02 program, the Belgian IAP, the Fund for Scientific
Research -- Flanders (F.W.O.--Vlaanderen), by the F.W.O. fellowship (A.V.S.), and by the ESF-NES Programme.  Work by C.J.O.R. and C.R. was carried out under
the NNSA of the US DoE at LANL under Contract No. DE-AC52-06NA25396.


\begin{thebibliography}{1}
\bibitem{comment}
For instance the earth magnetic field ($\sim$ 0.04 mT) if not properly shielded would give rise to an intervortex distance of about 7 $\mu$m.
\end{thebibliography}
\end{document}